\documentclass[showkeys,twocolumn,showpacs,preprintnumbers,amsmath,amssymb,aps,prb,a4paper]{revtex4}

\usepackage{graphicx}
\usepackage{bm}

\begin{document}

\pagenumbering{arabic}

\title{Reversible strain effect on the magnetization of LaCoO$_{3}$ films}

\author{A. Herklotz}
\author{A.~D. Rata}
\author{L. Schultz}
\author{K. D\"{o}rr}

\affiliation{IFW Dresden, Institute for Metallic Materials,
Helmholtzstra$\ss$e 20, 01069 Dresden, Germany }

\date{\today}

\begin{abstract}
The magnetization of ferromagnetic LaCoO$_3$ films grown
epitaxially on piezoelectric substrates has been found to
systematically decrease with the reduction of tensile strain. The
magnetization change induced by the reversible strain variation
reveals an increase of the Co magnetic moment with tensile strain.
The biaxial strain dependence of the Curie temperature is
estimated to be below $4$~K/$\%$ in the as-grown tensile strain
state of our films. This is in agreement with results from
statically strained films on various substrates.

\end{abstract}


 \maketitle



The perovskite-type LaCoO$_3$ has been intensively studied, mainly
because of the thermally driven spin state transitions of
Co$^{3+}$ ions, which give rise to unique properties
\cite{Imada,Goodenough71}. Despite large experimental and
theoretical efforts
\cite{Maris,Haverkort06,Podlesnyak06,Klie07,Korotin,Pandey08,Hozoi08},
the spin state of LaCoO$_3$ at finite temperatures remains
controversial. The ground state of LaCoO$_3$ is nonmagnetic with
Co$^{3+}$ ions in the low spin (LS,~$S\!=\!0$) state \cite{Imada}.
At temperatures above $100$~K, Co$^{3+}$ ions display various spin
states due to a delicate balance between the crystal-field
splitting $\Delta_{\mathrm{CF}}$ and the intraatomic Hund exchange
\cite{Imada,Raccah67}. Since $\Delta_{\mathrm{CF}}$ is very
sensitive to the variation of the Co-O bond length \cite{Sherman},
structural changes may easily modify the Co spin state. Recently,
LaCoO$_3$ has attracted renewed interest due to the observation of
ferromagnetism in epitaxially strained thin films
\cite{Fuchs07,Fuchs08}. Actually, the existence of either long- or
short-range ferromagnetic order has been reported for various
types of LaCoO$_3$ samples
\cite{Zhou07,Menyuk67,Yan04,Androulakis01,Harada07}. The origin of
the observed ferromagnetism is currently under investigation.

Thin films of complex magnetic oxides grown epitaxially on
piezoelectric single-crystalline substrates of
Pb(Mg$_{1/3}$Nb$_{2/3}$)$_{0.72}$Ti$_{0.28}$O$_3$(001)~(PMN-PT)
allow direct investigations of their strain-dependent magnetism
\cite{Thiele07,Rata08}. The as-grown strain state of the films can
be dynamically tuned by the inverse piezoelectric effect of the
substrate, thus providing an opportunity to record the change with
respect to strain of the ordering temperature or the magnetization
for the particular as-grown strain state. Thus, the influence of
strain on magnetism can be clearly separated from the effects of
other parameters, \textit{e.g.} oxygen nonstoichiometry, chemical
inhomogeneities, and microstructure.

We report on the influence of reversible strain on the
magnetization of thin LaCoO$_3$ films epitaxially grown on
PMN-PT(001). A roughly linear decrease of the magnetization with
the piezoelectrically controlled release of tensile strain is
observed at various temperatures below the Curie temperature
($T_{\mathrm{C}}$). From the measured reversible magnetization
change, an upper limit for the strain-dependent shift of
$T_{\mathrm{C}}$ is estimated. Strain-dependent magnetization data
give evidence for an enhanced magnetic moment of Co ions under
tensile strain.

LaCoO$_3$~(LCO) films of various thicknesses were grown on
PMN-PT(001) by pulsed laser deposition (KrF $248$~nm) from a
stoichiometric target. The deposition temperature ($T$) and the
oxygen background pressure were $650^{\circ}$C and $0.45$~mbar,
respectively. After deposition, the films were annealed for
$10$~minutes and cooled down in oxygen atmosphere of $800$~mbar.
Structure and film thickness were characterized by X-ray
diffraction (XRD) measurements with a Philips X'Pert MRD
diffractometer using Cu K$\alpha$ radiation. The magnetization
($M$) was measured in a SQUID magnetometer. $T_{\mathrm{C}}$ is
estimated by extrapolating $M^{2}$ for $T\!<\!T_{\mathrm{C}}$ to
$M\!=\!0$. For strain-dependent measurements, an electrical
voltage is supplied to the substrate between the magnetic film and
a bottom electrode on the opposite (001) surface of the substrate.
The current in the piezo-circuit is below $10^{-6}$~A.

X-ray $\Theta\!-\!2\Theta$ scans show clear (00l) film reflections
characteristic of a pseudocubic structure (Fig.~1). PMN-PT has a
pseudocubic lattice parameter of $4.02$~$\mathrm{\AA}$ and weak
rhombohedral or monoclinic distortions \cite{Thiele07}. Bulk LCO
is rhombohedral with a pseudocubic lattice parameter of
$3.805$~$\mathrm{\AA}$ \cite{Radaelli02}. Despite the large misfit
of $5.7$~$\%$, LCO grows epitaxially oriented on PMN-PT(001).
In-plane X-ray reciprocal space mapping around the (013)
asymmetric reflection reveals partial relaxation of the LCO film,
\textit{i.e.} the reflections of the film and the substrate have
different Q$_{x}$ values (Fig.~1 inset). The derived out-of-plane
(c) and in-plane (a) lattice parameters are
$c\!=\!3.79$~$\mathrm{\AA}$ and $a\!=\!3.88$~$\mathrm{\AA}$ for
the $50$~nm thick film. The tetragonal distortion estimated as $t
= c /a$ is $0.977$.
\begin{figure}[!t]
\includegraphics[width=7cm]{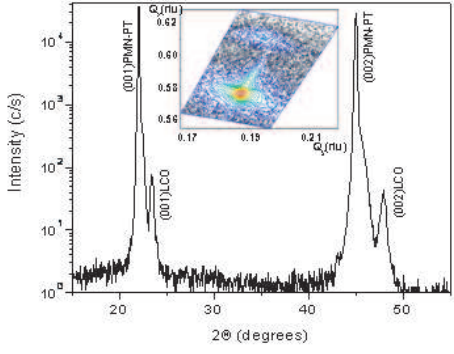}
\caption{$\Theta-2\Theta$ XRD of a $50$~nm thick LCO/PMN-PT(001)
film. Inset: XRD reciprocal space map around the (013) reflection.
The intense reflection corresponds to the substrate peak.}
\end{figure}
\begin{figure}
\includegraphics[width=7cm]{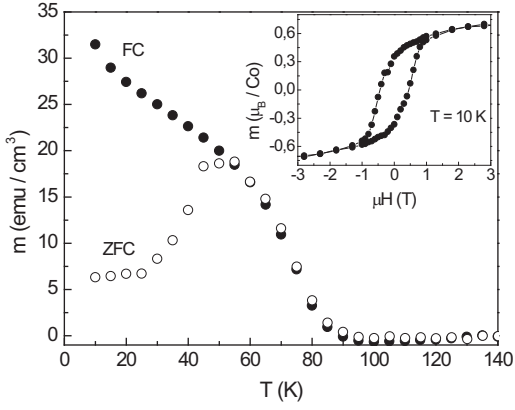}
\caption{Temperature-dependent in-plane magnetization of a $50$~nm
LCO film measured both in field-cooled and zero-field-cooled modes
in a field of $200$~mT along the [100] substrate direction. Inset:
magnetic hysteresis loop at $10$~K.}
\end{figure}

In Fig.~2 we plot $M$ vs $T$ of the film from Fig.~1 measured in
field-cooled (FC) and zero-field-cooled (ZFC) modes in a magnetic
field ($H$) of $\mu_{0}H\!=\!200$~mT applied along the in-plane
[100] direction. Magnetic ordering is observed below about $87$~K.
A cusp in the ZFC magnetization is found at $\sim50$~K, possibly
indicating a glass-like behaviour as in bulk cobaltites. The inset
shows the magnetization loop M(H) at $10$~K. Clearly, the LCO film
is ferromagnetic at low temperatures, with
$T_{\mathrm{C}}\simeq87$~K. This value agrees well with recently
published data for strained LCO films \cite{Fuchs07,Fuchs08}. The
coercive field ($H_c$) at $10$~K is $450$~mT.

\begin{figure}[!t]
\includegraphics[width=9cm]{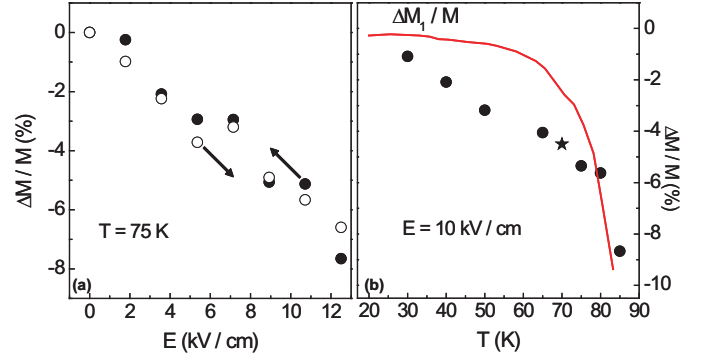}
\caption{(a) Electric field dependence of the strain-induced
magnetization ($M$) change calculated as $[M(E)-M(0)]/M(0)$ at
$T\!=\!75$ K. (b) Temperature dependence of the strain-induced $M$
change at $E\!=\!10$ kV/cm. Star data point was measured in
saturated state. All other data were recorded in FC mode at
$200$~mT. Solid line denotes the calculated $M$ change for
$\Delta$$T_\mathrm{C}=0.4$ K (see text).}
\end{figure}

\begin{table}[t]
\caption{In-plane ($a$) and out-of-plane ($c$) lattice parameters
of various LCO films grown under tensile strain, unit cell volume
($V$ = $a^2c$) and $T_\mathrm{C}$. $a_{sub}$ denotes the
pseudocubic substrate lattice parameter.}
\begin{ruledtabular}
\begin{tabular}{lccccc}
LSCO films     &$a_{sub}$ (\AA )
                           &$c$ (\AA )
                                        &$a$ (\AA )
                                                     &$V$( ${\mathrm{\AA}}^{3}$ )
                                                                 &$T_\mathrm{C}$ (K)  \\
(100 nm)   \\
\colrule
\\
LCO/LSAT       &3.87       &3.804         &3.867        &56.88       &85         \\
LCO/STO        &3.905       &3.785        &3.896        &57.45       &86         \\
LCO/PMN-PT    &4.02        &3.81          &3.87         &57.06       &87        \\
LSCO/LAO       &3.79           &3.85       &3.789        &55.27       &75         \\
\end{tabular}
\end{ruledtabular}
\end{table}

In the following we discuss the effect of reversible biaxial
strain on the magnetization of LCO. Fig.~3a shows the $M$
dependence on the applied substrate electric field ($E$) at
$75$~K, recorded in $200$~mT after field cooling. The increasing
electric field leads to linear in-plane compression of the
substrate \cite{Thiele07} and, hence, to reduction of the tensile
film strain. The value of the piezoelectrical substrate strain at
$10$~kV/cm is about $-0.1$~$\%$ at $90$~K \cite{Kathrin}. A
roughly linear, low-hysteresis decrease of $M$ is observed with
increasing $E$, \textit{i.e.} with decreasing tensile strain. A
similar behaviour occurs at various temperatures; the resulting
$M$ change measured at $E=10$~kV/cm is summarized in Fig.~3b. A
$M$ change of $\approx9\%$ is obtained near $T_\mathrm{C}$. In
order to clarify the strain effect also for the saturated
magnetization, full $M(H)$ loops ($H<5$~T) were recorded at $70$~K
in different strain states. They revealed a similar strain-induced
change of the saturated magnetization; see, \textit{e.g.}, the
data point for $70$~K inserted in Fig.~3b.

The observed decrease of $M$ with reduced tensile strain confirms
the observation of Fuchs \textit{et al.} \cite{Fuchs07,Fuchs08}
that tensile film strain stabilizes ferromagnetism in LaCoO$_3$.
This may involve strain-induced enhancements of (i) the magnetic
moment of the Co ions and (ii) $T_\mathrm{C}$ with tensile strain.
In the following we attempt to separate both effects.

As a first step, the $T_\mathrm{C}$ shift induced by the
reversible strain is estimated. The $M$ change
$\Delta$$M_{T_\mathrm{C}}$ resulting from a $T_\mathrm{C}$ shift
can be approximated by shifting the $M(T)$ curve (recorded under
equal conditions as for the reversible strain runs) by an assumed
temperature interval $\Delta$$T_\mathrm{C}$ and taking the
difference to the original data. The thus obtained $M$ change,
denoted as $\Delta$$M_{1}(T)$, overestimates
$\Delta$$M_{T_\mathrm{C}}$ at lower $T$ and converges to its real
value close to $T_\mathrm{C}$. Fig.~3b shows the
$\Delta$$M_{1}(T)$ curve calculated for
$\Delta$$T_\mathrm{C}=0.4$~K. This value of $\Delta$$T_\mathrm{C}$
is chosen to fit the measured $\Delta$$M$ at $80$~K, close to
$T_\mathrm{C}$. Therefore, it provides an upper limit for the real
$T_\mathrm{C}$ shift caused by the reversible strain of $0.1\%$ in
the LCO film. Hence, for the as-grown state of
$a\!=\!3.88$~$\mathrm{\AA}$ of the LCO film, the biaxial strain
change of the transition temperature is estimated as
$dT_\mathrm{C}/da \leq 4$~K/$\%$.

As a consequence of the above arguments, the measured
strain-induced $M$ change in the range of $T\!=\!30\div70$~K
cannot originate from a $T_\mathrm{C}$ shift alone, since it
substantially exceeds $\Delta$$M_{1}(T)$ giving an upper limit to
the $M$ change caused by the $T_\mathrm{C}$ shift. Clearly, a
decrease of the Co magnetic moment itself is needed to explain the
data.

The above estimated maximum shift of $T_\mathrm{C}$ of $4$~K/$\%$
of biaxial strain for the as-grown state of $a(\mathrm{LCO}) =
3.88$~$\mathrm{\AA}$ can be compared to the Curie temperature vs
strain for LCO films grown on various substrates, see Tab.~1 and
Ref.~\cite{Fuchs08}. We find little variation of $T_\mathrm{C}$ in
the range of $2$~K for $a =3.867 \div 3.896$ $\mathrm{\AA}$,
indicating a weaker than the estimated maximum strain response of
$T_\mathrm{C}$. It is worth noting that a compressively strained
LCO film ($a\!=\!3.789$~$\mathrm{\AA}$) grown on LaAlO$_3$ shows a
remanent magnetization up to $75$~K, too. Hence, the strain effect
on $T_\mathrm{C}$ appears to be rather moderate.

Finally, it is interesting to consider the roles played by (i) the
tetragonal distortion of the film characterized by the $c/a$ ratio
and (ii) the strain-induced volume ($V$) change of the unit cell.
Tensile strain typically increases $V$. Thus, its effect is
opposite to hydrostatic pressure. Co ions in LaCoO$_3$ have been
reported to transfer to the low-spin state under hydrostatic
pressure \cite{Vanko06}, as is also found for the doped
La$_{0.82}$Sr$_{0.18}$CoO$_3$ \cite{Lengsdorf07}. This is
consistent with the enlarged ionic radius of Co ions in the
excited, \textit{i.e.} intermediate/high, spin states
\cite{Radaelli02}. Hence, a $V$ increase is likely to stabilize
excited spin states of Co ions. Zhou \textit{et al.} \cite{Zhou07}
indicated that volume increase may underlie the ferromagnetism
observed in the LCO nanoparticles investigated in their work. The
tetragonal distortion, on the other hand, seems to be less
important for establishing ferromagnetism in LCO, (even though it
may have an influence), since a high $T_\mathrm{C}$ of about
$85$~K has been reported for samples without tetragonal
distortion, \textit{e.g.} for nanoparticles \cite{Zhou07} and the
LCO/SrLaAlO$_4$ film discussed in Ref.~\cite{Fuchs08}.

Summarizing, the influence of reversible biaxial strain on the
magnetization of epitaxially grown LaCoO$_3$ films has been
investigated. The strain-induced increase of T$_C$ is estimated to
be below $4$~K/\% of strain in the as-grown state of
$a\!=\!3.88$~$\AA$. Our data give evidence for an enhanced
magnetic moment of Co ions under tensile strain. Both results
confirm that tensile strain strengthens the ferromagnetism in
LaCoO$_3$ films. Dominance of the effect of an enlarged unit cell
volume over that of a tetragonal distortion is suggested for
inducing ferromagnetism by tensile strain. Soft X-ray absorption
experiments, which may clarify the effect of strain on the
electronic structure, and particularly on the Co spin state, are
in progress.

We thank R. H\"{u}hne for stimulating discussions. This work was
supported by Deutsche Forschungsgemeinschaft, FOR $520$.

\end{document}